\documentclass[aps,prl,twocolumn,superscriptaddress]{revtex4-1}

\usepackage{graphicx}
\usepackage{dcolumn}
\usepackage{bm}
\usepackage{hyperref}
\usepackage{todonotes}
\usepackage{color}
\usepackage{comment}
\usepackage{sidecap}
\sidecaptionvpos{figure}{c}

\begin{document}

\title{Orbital engineering in nickelate heterostructures driven by anisotropic oxygen hybridization rather than orbital energy levels}

\author{G. Fabbris}
\email{gfabbris@bnl.gov}
\author{D. Meyers}
\affiliation{Department of Condensed Matter Physics and Materials Science, Brookhaven National Laboratory, Upton, New York 11973, USA}

\author{J. Okamoto}
\affiliation{National Synchrotron Radiation Research Center, Hsinchu 30076, Taiwan}

\author{J. Pelliciari}
\affiliation{Research Department ``Synchrotron Radiation and Nanotechnology'', Paul Scherrer Institut, CH-5232 Villigen PSI, Switzerland}

\author{A. S. Disa}
\affiliation{Department of Applied Physics, Yale University, New Haven, Connecticut 06520, USA}

\author{Y. Huang}
\affiliation{Research Department ``Synchrotron Radiation and Nanotechnology'', Paul Scherrer Institut, CH-5232 Villigen PSI, Switzerland}

\author{Z.-Y. Chen}
\author{W. B. Wu}
\author{C. T. Chen}
\affiliation{National Synchrotron Radiation Research Center, Hsinchu 30076, Taiwan}

\author{S. Ismail-Beigi}
\affiliation{Department of Applied Physics, Yale University, New Haven, Connecticut 06520, USA}

\author{C. H. Ahn}
\affiliation{Department of Applied Physics, Yale University, New Haven, Connecticut 06520, USA}
\affiliation{Department of Mechanical Engineering and Materials Science, Yale University, New Haven, Connecticut 06520, USA}

\author{F. J. Walker}
\affiliation{Department of Applied Physics, Yale University, New Haven, Connecticut 06520, USA}

\author{D. J. Huang}
\affiliation{National Synchrotron Radiation Research Center, Hsinchu 30076, Taiwan}
\affiliation{Department of Physics, National Tsing Hua University, Hsinchu 30013, Taiwan}

\author{T. Schmitt}
\affiliation{Research Department ``Synchrotron Radiation and Nanotechnology'', Paul Scherrer Institut, CH-5232 Villigen PSI, Switzerland}

\author{M. P. M. Dean}
\email{mdean@bnl.gov}
\affiliation{Department of Condensed Matter Physics and Materials Science, Brookhaven National Laboratory, Upton, New York 11973, USA}

\def\mathbi#1{\ensuremath{\textbf{\em #1}}}
\def\Q{\ensuremath{\mathbi{Q}}}
\def\LNO{LaNiO$_3$}
\newcommand{\angstrom}{\mbox{\normalfont\AA}}
\date{\today}

\begin{abstract}
Resonant inelastic x-ray scattering is used to investigate the electronic origin of orbital polarization in nickelate heterostructures taking $\mathrm{LaTiO_3-LaNiO_3-3x(LaAlO_3)}$, a system with exceptionally large polarization, as a model system. We find that heterostructuring generates only minor changes in the Ni $3d$ orbital energy levels, contradicting the often-invoked picture in which changes in orbital energy levels generate orbital polarization. Instead, O $K$-edge x-ray absorption spectroscopy demonstrates that orbital polarization is caused by an anisotropic reconstruction of the oxygen ligand hole states. This provides an explanation for the limited success of theoretical predictions based on tuning orbital energy levels and implies that future theories should focus on anisotropic hybridization as the most effective means to drive large changes in electronic structure and realize novel emergent phenomena. 
\end{abstract}


\maketitle

\begin{figure}
\includegraphics{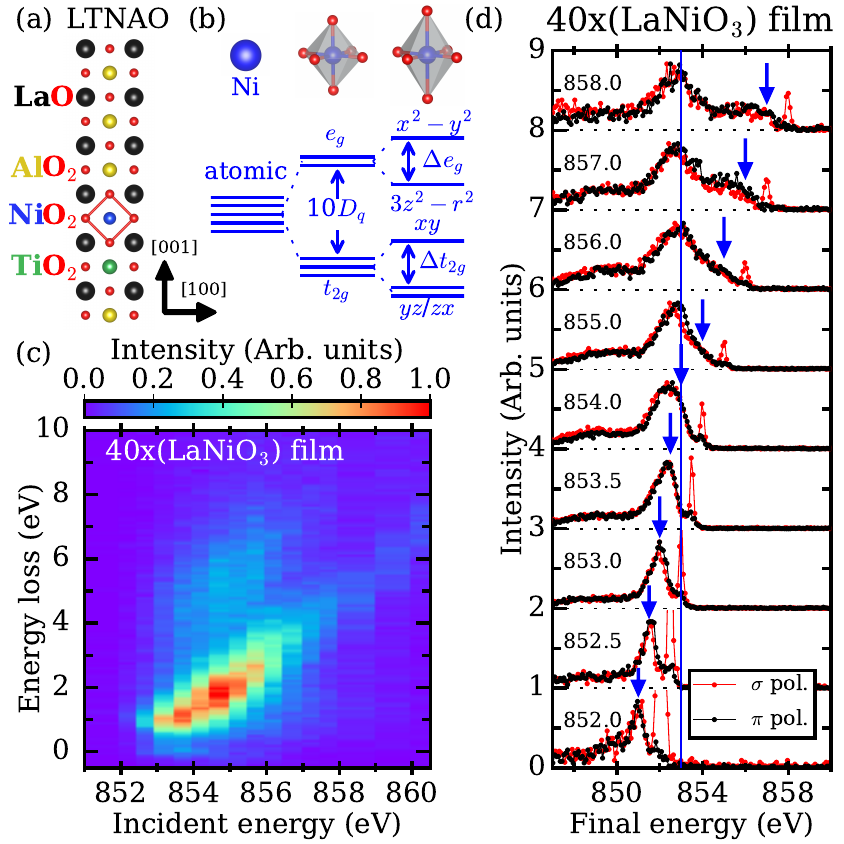}
\caption{(Color online) (a) The basic structural unit studied here composed of $\mathrm{LaTiO_3}$, $\mathrm{LaNiO_3}$ and $\mathrm{LaAlO_3}$ layers in a 1:1:3 ratio. (b) Energy level diagram for the Ni electronic state. (c) Ni $L_3$-edge RIXS map of a LaNiO$_3$ film measured with $\pi$-polarized incident x-rays.  The strongest inelastic spectral feature is the diagonal constant final energy emission line characteristic of a metallic system. (d) Normalized RIXS spectra as a function of final energy, for different incident energies and both polarizations as labeled. A vertical blue line at 853~eV marks the fluorescence feature. Blue arrows denote 1.1~eV energy loss. The sharp feature observed at larger final energies for $\sigma$ is the elastic line. For incident energies 852-853.5~eV and 857-859~eV a 1.1~eV constant energy loss feature is visible.}
\label{LNO_struct_map_linecuts}
\end{figure}

The electronic structure of transition metal oxides (TMOs) is dominated by the active $3d$ TM orbitals and how these hybridize with the neighboring oxygen $2p$ ligand orbitals. Building heterostructures from one-unit-cell-thick layers of different TMOs offers the opportunity of tuning the TM $3d$ states configuration and potentially realizing emergent phenomena with new or improved properties \cite{Schlom2008, Zubko2011, Hwang2012, Bhattacharya2014, Chakhalian2014, Middey2016}. \LNO{} based heterostructures are a prototypical example of such an endeavor \cite{Benckiser2011, Chakhalian2011, Freeland2011, Han2010, Han2011, Liu2011, Gibert2012, Wu2013, Chen2013, Kumah2014, Disa2015, Disa2015_extra, Cao2016, Grisolia2016, Middey2016}, motivated by several predictions of novel superconducting, magnetic and topological states \cite{Chaloupka2008, Hansmann2009, Ruegg2011, Yang2011b}. Bulk \LNO{} has \emph{nominally} Ni$^{3+}$ ions with filled $t_{2g}$ states and one $e_g$ electron. Efforts to realize the predicted novel states are crucially dependent on inducing orbital polarization in the Ni $e_g$ level \footnote{For the purposes of this paper, we define orbital polarization in terms of the presence of strong x-ray linear dichroism as explained further in the supplemental material \cite{supplemental}}, i.e. breaking its degeneracy and pushing the system towards a half filled $x^2-y^2$ (or $3z^2-r^2$) configuration (Fig. \ref{LNO_struct_map_linecuts}(b)). Initial efforts to realize strong orbital polarization in heterostructures reported up to $\sim20$\% change in orbital occupancy compared to the bulk \cite{Benckiser2011, Chakhalian2011, Freeland2011, Han2010, Han2011, Liu2011, Wu2013, Cao2016}, corresponding to $\sim$ 0.3 eV splitting of the $e_g$ states as inferred from cluster calculations of x-ray absorption spectra \cite{Wu2013}, in fact first principle calculations indicate that $\sim 10$\% strain is needed to drive $\sim 40$\% change in orbital polarization (see supplemental material of Ref. \onlinecite{Disa2015}). Novel trilayer $\mathrm{LaTiO_3-LaNiO_3-3x(LaAlO_3)}$ (LTNAO) superlattices were recently produced obtaining $\sim 50$\% change in orbital polarization via charge transfer, polar charge and electronic confinement effects \cite{Chen2013, Kumah2014, Disa2015, Disa2015_extra}. To date, changes in orbital polarization have been conceptualized as a consequence of tuning the 3$d$ orbital energy level through crystal field engineering. However, this approach has led to a dramatic mismatch between theoretical predictions and experimental results \cite{Chaloupka2008, Hansmann2009,Ruegg2011,Yang2011,Wu2013}, raising questions about the precise electronic character, and the driving mechanism behind, orbital polarization in these materials.


We apply resonant inelastic x-ray scattering (RIXS) \cite{Ament2011, Dean2015} to $\mathrm{LaNiO_3}$ and LTNAO superlattices obtaining a far more detailed spectral fingerprint of the Ni electronic configuration than is possible with standard x-ray absorption spectroscopy (XAS) \cite{Chakhalian2011, Benckiser2011, Chakhalian2011, Liu2011, Freeland2011, Wu2013, Disa2015, Cao2016,MeyersENO2013}. Charge transfer is found to drive \LNO{} out of its initial itinerant state in the bulk into a more localized state in LTNAO superlattices that display Ni $d^8 \underline{L}$ electronic configuration (where $\underline{L}$ denotes partial oxygen ligand hole character). A crystal field splitting $\Delta e_g$ = 0.20(5)~eV is found in LTNAO superlattices, far smaller than previously thought and inconsistent with the often invoked picture of orbital polarization driven by crystal field splitting \cite{Chaloupka2008, Hansmann2009, Chakhalian2011, Wu2013, Chen2013, Disa2015}. X-ray absorption spectroscopy at the O $K$-edge demonstrates that orbital polarization is instead driven by anisotropic hybridization with the oxygen ligands. This explains the current lack of success in realizing novel emergent phenomena in nickelate heterostructures based on changing orbital energy levels \cite{Chaloupka2008, Hansmann2009, Ruegg2011, Yang2011b}. Our work indicates that manipulating the anisotropy of the Ni hybridization with oxygen will be the most effective route to drive large changes in electronic structure of nickelates and therefore provides the best chances of realizing novel emergent states. 

Samples were synthesized on LaAlO$_3$ [001] substrates using oxygen plasma-assisted molecular beam epitaxy (MBE) as described in detail in Ref.~\onlinecite{Disa2015}. The LTNAO sample is depicted in Fig.~\ref{LNO_struct_map_linecuts}(a) and was prepared with a layering sequence {$\rm [LaTiO_3-LaNiO_3-3x(LaAlO_3)]  \times 3$}. The \LNO{} sample was a 40 formula unit thick reference. The $\mathrm{LaAlO_3}$ substrate introduces a $\sim 1.1$\% compressive strain in $\mathrm{LaNiO_3}$. X-ray characterization shows excellent structural quality with roughnesses of approximately 5~\AA\ \cite{supplemental}. We determined the dichroism in the XAS signal of LTNAO by integrating our RIXS data along the energy loss axis and found $\sim 50$\% increase in the in-plane character of the orbital polarization \cite{supplemental} consistent with previous work \cite{Disa2015}.

During the RIXS experiments, a $(H,0,L)$ horizontal scattering plane was employed. X-rays were incident at $15^\circ$ with respect to the sample surface polarized either parallel ($\pi$) or perpendicular ($\sigma$) to the scattering plane and x-rays scattered around $90^{\circ}$ were energy analyzed by the spectrometer \cite{supplemental}. The LTNAO sample was measured with the SAXES spectrometer \cite{Ghiringhelli2006} located at the ADRESS beamline \cite{Strocov2010} of the Swiss Light Source and the LaNiO$_3$ sample was measured with the AGS-AGM setup \cite{Lai2014} at BL05A1 - the Inelastic Scattering Beamline at the National Synchrotron Radiation Research Center, Taiwan. The combined energy-loss resolution for both experiments was $\sim 210$~meV.  All data were collected at $\sim13$~K to reduce thermal diffuse scattering.

When grown in the bulk, LaNiO$_3$, the active constituent of the heterostructures studied here, is a correlated metal with approximately cubic symmetry \cite{Catalan2008}. The \emph{nominal} Ni$^{3+}$ ground state is usually interpreted as a mixture between $d^7$ and $d^8 \underline{L}$ \cite{Medarde1997, Mizokawa1995, Piamonteze2005, Liu2011, Johnson2014, Freeland2016, Cao2016, Bisogni2016}. Figure \ref{LNO_struct_map_linecuts}(c) plots a map of the RIXS intensity of the $\mathrm{LaNiO_3}$ film. The main spectral feature appears as a diagonal, constant final energy emission line, that is visible for incident energies above the Ni $L_3$ resonance around 853~eV, as is typically observed for systems with mostly metallic character \cite{Jimenez-Mier1999, Finkelstein1999, Kotani2001}. Figure \ref{LNO_struct_map_linecuts}(d) plots individual spectra as a function of final energy, normalized to peak intensity, which reveals a weak Raman-like constant energy loss feature at $\sim$~1.1~eV, suggesting a partially localized component to the electronic character.  Recent RIXS and theoretical work place \LNO{} in the negative charge transfer regime \cite{Bisogni2016, Anisimov1999, Mizokawa2000, Park2012, Johnston2014,Parragh2013,  Nowadnick2015,Park2016}, with RIXS Raman features coming from 3$d^8\underline{L}$ states. The polarization dependence of the $\sim 1.1$~eV feature is directly related to the $e_g$ splitting \cite{supplemental}. Therefore, the weak polarization dependence shown in Fig.~\ref{LNO_struct_map_linecuts}(d) places an upper limit in the strain-induced $e_g$ splitting $\Delta e_g \lesssim 0.2$~eV \cite{supplemental} in  agreement with the literature \cite{Benckiser2011, Chakhalian2011, Freeland2011, Han2010, Han2011, Liu2011, Wu2013}.

\begin{figure}
\includegraphics[width=3.in]{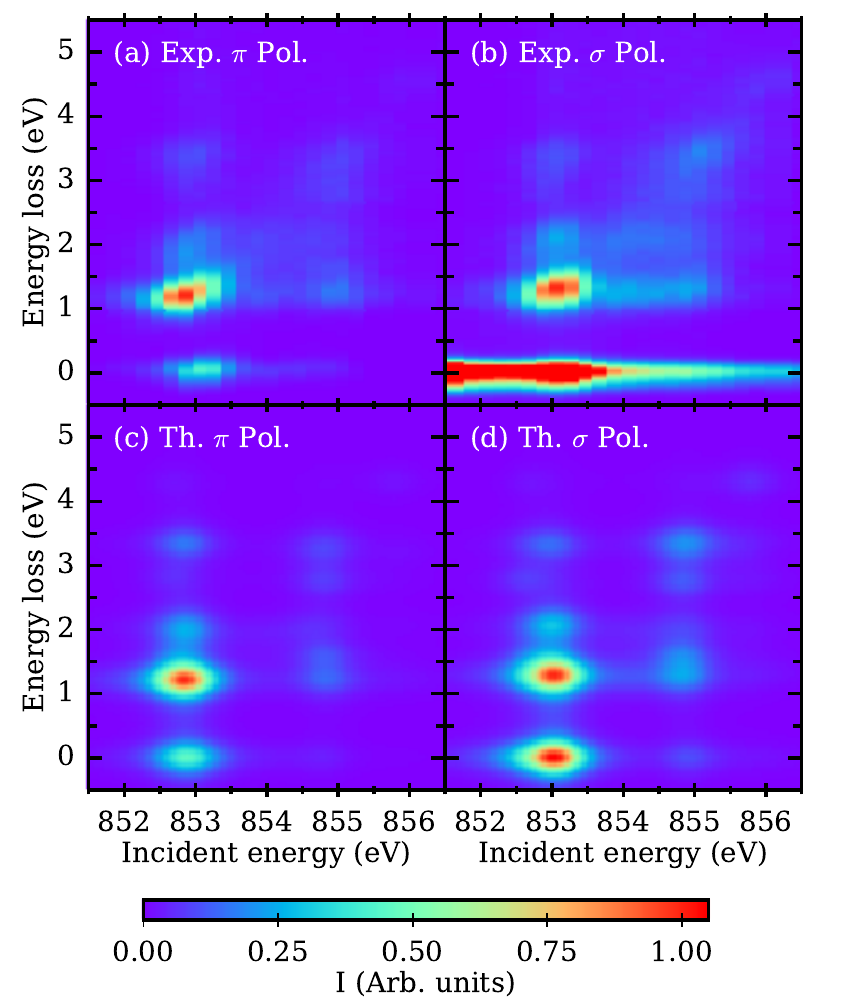} %
\caption{(Color online) Color maps of the RIXS intensity of the LTNAO heterostructure. The spectra are dominated by Raman-like constant energy loss excitations that resonant differently as a function of incident energy, providing a detailed spectral fingerprint of the Ni electronic structure in the heterostructure. (a)\&(b) plot the measured intensity with $\pi$ and $\sigma$ polarized incident x-rays. The strong elastic peak at low incident energies for $\sigma$ polarization is due to the La $M_4$ edge. (c)\&(d) plot our corresponding multiplet calculations for the two different polarizations.}
\label{RIXS_maps}
\end{figure}

\begin{figure}
\includegraphics[width=3.in]{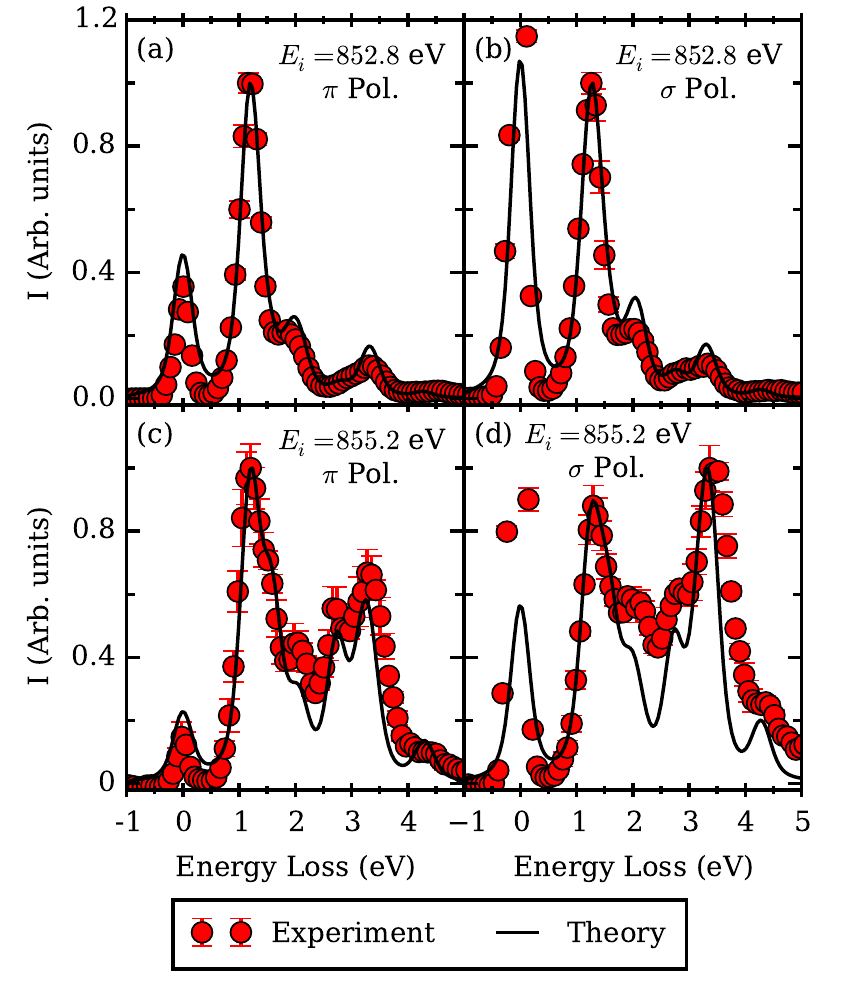} %
\caption{(Color online) Comparison of the experimental and theoretical RIXS spectra of the LTNAO heterostructure at fixed incident energy. Peak energies are reproduced within about 0.1~eV and intensities are reproduced within $\sim 20$\%. (a)\&(b) plot spectra at an incident energy of $E_i=852.8$~eV for $\pi$ and $\sigma$ polarized incident x-rays, respectively. (c)\&(d) plot the same quantities for $E_i=855.2$~eV. Note that for $\sigma$ incident x-rays the elastic line intensity is a combination of RIXS intensity and diffuse scattering not included in the model, so full agreement is not necessarily expected.}
\label{line_cuts}
\end{figure}

Figure \ref{RIXS_maps}(a)\&(b) plots the dramatic change in the Ni $L_3$ edge RIXS maps once \LNO{} is confined within a LTNAO heterostructure. Rather than constant final energy fluorescence, the spectra now feature Raman-like constant energy loss excitations that resonate differently as a function of incident energy. In this regard the Ni electronic state looks much more similar to correlated insulators with a Ni $d^8$ configuration such as NiO or NiCl$_2$ \cite{Magnuson2002, Ghiringhelli2005, Ghiringhelli2009}. Indeed, LTNAO has been observed to be far more electrically insulating than \LNO{} \cite{Disa2015}.  By measuring both the incident and scattered x-ray photon energies, RIXS obtains many more distinct spectral features than XAS, particularly because the linewidth of the RIXS final states are not limited by the core hole lifetime. We analyzed our data using multiplet calculations in the Cowan-Butler-Thole approach \cite{deGroot2005, deGrootBook, Stavitski2010}, which computes the $L$-edge RIXS signal for an ion within a given crystal field. These methods are widely used to model resonant x-ray spectra of TMOs \cite{deGrootBook} and are explained in more detail in the Supplemental Material \cite{supplemental}.

We found that the LTNAO RIXS can be modelled in terms of $dd$ excitations from a Ni $d^8$ atom in a tetragonal crystal field as plotted in Fig.~\ref{RIXS_maps} with linecuts shown in Fig.~\ref{line_cuts}. The parameters describing the LTNAO electronic configuration are detailed in Table \ref{CTM_parameters}, and Fig.~\ref{LNO_struct_map_linecuts}(b) illustrates the parameters $10D_q$, $\Delta t_{2g}$ and $\Delta e_g$ in terms of the orbital energy levels. Details on the estimated errors are described in the Supplemental Material \cite{supplemental}. The atomic model used here assigns only $d$ states to the valence band, hence the extracted energies correspond to those of effective hybridized orbitals in the electronic structure. We find excellent agreement between experiment and theory for the peak energies, as good, or better, than multiplet calculations for much simpler bulk compounds \cite{Magnuson2002, Ghiringhelli2005, Ghiringhelli2009}. A slightly poorer agreement is found regarding the intensity of the excitations, which is likely due to small intensity renormalization driven by the presence of ligand holes \cite{Ghiringhelli2005}. Nevertheless, RIXS shows that heterostructuring has driven the system closer to a Ni $d^8$ electronic configuration than a Ni $d^7$ configuration ($d^7$ calculations completely fail to reproduce our measurements \cite{supplemental}).
 
\begin{table}[b]
\begin{ruledtabular}
\caption{The parameters describing the electronic configuration of Ni in the LTNAO heterostructure. $10D_q$, $\Delta t_{2g}$ and $\Delta e_g$ describe the crystal field energy levels (see Fig.~\ref{LNO_struct_map_linecuts}(b)) and $F$ and $G$ control the values of the Slater-Condon parameters \cite{deGrootBook}. $F_{dd}$ and $F_{pd}$ are the percentage reductions in the Coulomb repulsion between two $d$ electrons and $p$-$d$ electrons respectively. $G_{pd}$ is the percentage reduction in the Coulomb exchange. Errors were calculated as described in the Supplemental Material \cite{supplemental}}
\begin{tabular}{cccccccc}
\label{CTM_parameters}
  & $10D_q$ & $\Delta t_{2g}$ & $\Delta e_g$ & $F_{dd}$ & $F_{pd}$ & $G_{pd}$ \\
\hline
Values   & 1.28(3)~eV    & 0.05(5)~eV                & 0.20(5)~eV               & 65\%     & 95\%     & 70\% 
\end{tabular}
\end{ruledtabular}
\end{table}

The predominant Ni $d^8$ electronic configuration is likely to arise primarily from charge transfer effects, as unlike some other perovskites, the $3d$ levels in LaTiO$_3$  lie near to the Fermi level of \LNO{}-heterostructures \cite{Chen2013_calcs, Han2014}. Ti $L$-edge XAS showed that the Ti charge state changes from $3+$ to close to $4+$ upon heterostructuring indicating that a large fraction of the one electron per Ti has been transferred to other layers in the heterostructure \cite{Disa2015} driving Ni towards a $d^8$ state.

Orbital polarization in nickelate heterostructures is usually interpreted to be due to enhanced $\Delta e_g$ driven by tetragonal distortions \cite{Chaloupka2008, Hansmann2009, Chakhalian2011, Wu2013, Chen2013, Disa2015}. The ratio between out- and in- plane Ni-O bond length is as large as 1.16 in LTNAO \cite{Disa2015}, an anisotropy comparable to that of La$_2$NiO$_4$, for which $\Delta e_g \sim 0.7$~eV \cite{Kuiper1998}. Indeed,  $\Delta e_g \sim 0.8$~eV was suggested previously by projecting the DFT band structure onto effective hybridized  $3z^2 - r^2$ and $x^2 - y^2$ Wannier orbitals \cite{Disa2015}. However, the effective Ni crystal field levels measured by RIXS in LTNAO yields $\Delta e_g = 0.20(5)$~eV, at most $\sim$~0.1~eV larger than what is induced by strain alone. The large reconstruction of oxygen ligand holes (discussed below) may explain such unexpectedly weak structural dependence of $\Delta e_g$, but further work is needed to address this result. Nevertheless, our work demonstrates the ability of Ni $L_3$-edge RIXS in obtaining a precise description of the crystal field energies acting upon $3d$ orbitals in heterostructures. Coupled with ongoing efforts to improve theoretical calculations \cite{Vernay2008, Stavitski2010, Ament2011, Hozoi2011, Wray2012, Haverkort2014, Fernandez-Rodriguez2015}, this emphasizes the as yet underutilized potential of this technique to understand the properties of transition metal oxide heterostructures.


\begin{figure}
\includegraphics[width=3.in]{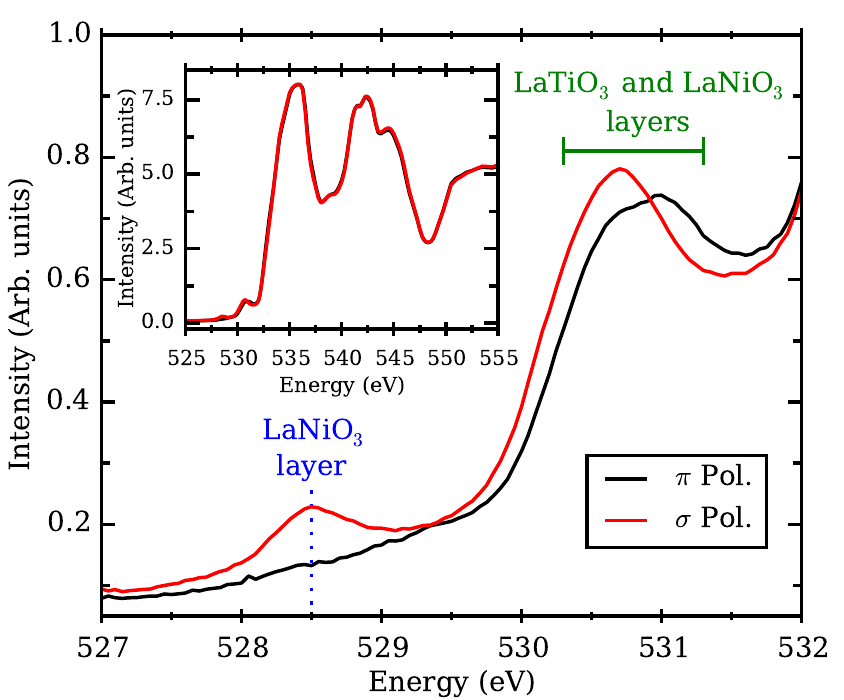}
\caption{(Color online) LTNAO O $K$-edge x-ray absorption spectra measured in total fluorescence yield mode. The inset shows the full energy window, the pre-edge region of which is plotted in the main panel.}
\label{okxas}
\end{figure}

In the face of LTNAO's small tetragonal crystal field splitting, the origin of Ni's orbital polarization needs to be revisited. Although Ni $L$-edge RIXS provides an accurate picture of orbital energy levels, it is much less sensitive to the degree of oxygen ligand hole character. In particular, the multiplet calculations used here yield a Ni $d^8$ high spin state, with formally one electron in the $3z^2 - r^2$ and $x^2 - y^2$ orbitals, and cannot explain the observed orbital polarization. The multiplet calculations, however, do not include the possibility of oxygen ligand holes, which can modify the effective orbital occupation. We therefore examined the oxygen ligand hole states by O $K$-edge XAS plotted in Fig.~\ref{okxas}. Every layer of the sample contains oxygen atoms that contribute to the main absorption edge and to features above the edge. On the other hand, pre-edge peaks that provide information about hybridization can be distinguished by comparing their energies to the literature \cite{Kuiper1995, Pellegrin1996, Suntivich2014, Cao2016}. The pre-peak at $\approx 528.5$~eV is assigned to Ni $3d^8 \underline{L}$ states in the LaNiO$_3$ layers; while the  $\approx 531$~eV features come from Ni $3d^9 \underline{L}$ states and Ti states in the LaTiO$_3$ layers \cite{Cao2016}. Thus Fig.~\ref{okxas} indicates that the Ni $d^8$ ground state features partial  Ni $d^8 \underline{L}$ character, likely coming from an incomplete Ti$\to$Ni charge transfer. Ligand holes appear to only weakly alter Ni $L_3$-edge RIXS, consistent with Anderson impurity model calculations for NiO \cite{Ghiringhelli2005}.

The intensity of the 528.5~eV LaNiO$_3$ pre-peak is proportional to the number of ligand holes \cite{Chen1992, Pellegrin1993, Kuiper1995, Pellegrin1996, Suntivich2014}. Remarkably, a very large linear dichroism is observed in this pre-peak (Fig.~\ref{okxas}), the $\pi$ channel is at least 20$\times$ weaker than $\sigma$ (in the former any signal is within the measurement error). This large anisotropy translates into more holes in the 3$d _{x^2-y^2}p_{\sigma}$ hybridized state than in the 3$d _{3z^2-r^2}p_{z}$ state, where $p_{\sigma}$ and $p_{z}$ denote in-plane and out-of-plane oxygen states, establishing that the large orbital polarization measured at the Ni edge originates from an uniquely anisotropic 3$d^8 \underline{L}$ state. The ligand hole anisotropy in LTNAO (as measured by XAS) is much closer to that observed in cuprates ($\sim10-100\times$) \cite{Chen1992,Pellegrin1993}, than in nickelates ($\sim1-2\times$) \cite{Kuiper1995,Pellegrin1996}. Therefore, heterostructuring successfully induced a dominating 3$d _{x^2-y^2}p_{\sigma}$ character to \LNO{} Fermi surface. Such character is among the defining properties of cuprate superconductors, thus placing LTNAO as the closest $S=1$ analogue to cuprates known to date. Probing the fermiology of LTNAO, particularly upon hole doping, remains as an outstanding opportunity for future investigations. 

In conclusion, we demonstrate that the large orbital polarization observed in LTNAO is driven by an anisotropic hybridization of the $3d-2p$ orbitals. This result largely disagrees with the common attempt to manipulate the properties of nickelate heterostructures by tuning the crystal field symmetry, an approach that has led to exciting predictions, but which have failed to materialize \cite{Chaloupka2008, Hansmann2009,Ruegg2011,Yang2011,Wu2013}. On the other hand, our results are congruent with the growing realization that \LNO{} resides in the negative charge transfer regime, dominated by the $3d^8\underline{L}$ configuration \cite{Bisogni2016, Anisimov1999, Mizokawa2000, Park2012, Parragh2013, Johnston2014, Grisolia2016}. The $3d^8$ high spin triplet is very stable. Due to the large Hund's and on-site Coulomb energies, a very large tetragonal splitting, of the order of 10$D_q$ (1.28(3) eV in LTNAO), is required to destabilize this ground state. A similar situation is realized in La$_2$NiO$_{4+\delta}$ ($\delta =$ 0, 0.05, 0.12) which has negligible ($<5$\%) orbital polarization  despite its very large $\Delta e_g \sim 0.7$~eV \cite{Kuiper1998}. Ni $L$-edge x-ray linear dichroism measurements of heavily doped La$_{2-x}$Sr$_x$NiO$_4$ quantifying possible orbital polarization in this system and comparing it to LTNAO would also be a highly interesting reference.  

Inducing substantial orbital polarization in the Ni 3$d$ state appears to be only achievable through anisotropic Ni-O hybridization. This implies that the origin of the orbital polarization in various nickelate heterostructures needs to be re-examined. Furthermore, a number of emergent phenomena occur in systems located in the small or negative charge transfer regime of the Zaanen-Sawatzky-Allen phase diagram \cite{Zaanen1985, Mizokawa1991,Khomskii2001}, in which ligand holes are the dominant state at the Fermi surface. Examples of such systems include CrO$_2$ \cite{Korotin1998}, BaFeO$_3$ \cite{Tsuyama2015}, SrCoO$_3$ \cite{Kunes2012}, and cuprates \cite{Mizokawa1991}. For instance, in cuprates the ligand hole couples to the 3$d$ hole to form a Zhang-Rice singlet \cite{Zhang1988b}, a state that is believed to be crucial for high-$T_c$ superconductivity. Therefore the ability to control the TM 3$d$ - O 2$p$ hybridization through charge transfer, polar charge, and electronic confinement effects, in tandem with first principles calculations that treat strong correlations and oxygen hybridization accurately such as dynamical mean field theory \cite{Park2016}, represents a highly promising route towards achieving the as yet largely unrealized aim of generating novel emergent states in transition metal oxide heterostructures. 

\begin{acknowledgments}
This material is based upon work supported by the U.S.\ Department of Energy, Office of Basic Energy Sciences, Early Career Award Program under Award Number 1047478. We thank  Frank de Groot, Valentina Bisogni, Liviu Hozoi, Jeroen van den Brink, Eli Stavitski, Jian Liu and John Tranquada for helpful discussions. J.P. and T.S. acknowledge financial support through the Dysenos AG by Kabelwerke Brugg AG Holding, Fachhochschule Nordwestschweiz, and the Paul Scherrer Institut. We also acknowledge support from the AFOSR and the NSF under MRSEC DMR-1119826 (CRISP). Experiments were performed at the ADRESS beamline of the Swiss Light Source at the Paul Scherrer Institut, Switzerland and at BL05A1 - Inelastic Scattering at National Synchrotron Radiation Research Center, Taiwan.
\end{acknowledgments}

\bibliography{refs}

\end{document}